\documentclass[11pt,a4paper]{article}

\usepackage{amsmath,amsthm,amssymb,amsfonts}
\usepackage{mathrsfs}
\usepackage{authblk}
\usepackage{setspace}
\usepackage{mathrsfs}
\usepackage{cases}
\usepackage{graphicx}
\usepackage{amsthm}
\usepackage[usenames]{xcolor}

\numberwithin{equation}{section}
\newcommand*{\M}{M_{\star}}

\newcommand*{\Mtilde}{{\tilde M}}
\newcommand*{\Etilde}{{\tilde E}}

\newcommand{\BbbR}{\mathbb{R}}
\newcommand{\BbbZ}{\mathbb{Z}}

\renewcommand*{\k}{\mathbf{k}}
\newcommand*{\x}{\mathbf{x}}

\newcommand*{\z}{\mathbf{z}}

\newcommand{\Finnormal}{\mathcal{F}^{in}_0}
\newcommand{\Fcircnormal}{\mathcal{F}^{circ}_0}

\DeclareMathOperator{\arctanh}{artanh}
\DeclareMathOperator{\arcsec}{arcsec}

\DeclareMathOperator{\Ai}{Ai}

\theoremstyle{plain}
\newtheorem{thm}{Theorem}[section]

\begin{document}

\title{Low energy Lorentz violation from high energy\\
modified dispersion in inertial and circular motion}
\author{Jorma Louko\footnote{jorma.louko@nottingham.ac.uk}\ }
\author{Samuel D. Upton\footnote{eeysdu@exmail.nottingham.ac.uk}}

\affil{School of Mathematical Sciences, University of Nottingham,\\
Nottingham NG7 2RD, UK}

\date{Revised December 2017\\[1ex]
Published in Phys.\ Rev.\ D {\bf 97}, 025008 (2018)}

\maketitle

\begin{abstract}
We consider an 
Unruh-DeWitt detector in inertial and circular motion in Minkowski spacetime 
of arbitrary dimension, 
coupled to a quantised scalar field with the Lorentz-violating dispersion relation 
$\omega = |\k|\, f\bigl(|\k|/\M\bigr)$, where $\M$ is the Lorentz-breaking scale. 
Assuming that $f$ dips below unity somewhere, we show that an inertial detector 
experiences large low energy Lorentz violations 
in all spacetime dimensions greater than two, 
generalising previous results in four dimensions. 
For a detector in circular motion, we show that a similar low energy 
Lorentz violation occurs in three spacetime dimensions, 
and we lay the analytic groundwork for examining 
circular motion in all dimensions greater than three, 
generalising previous work by  
Stargen, Kajuri and Sriramkumar 
in four dimensions. 
The circular motion results may be relevant for the prospects of observing  
the circular motion Unruh effect in analogue laboratory systems.  
\end{abstract}

\singlespacing

\section{Introduction\label{sec:intro}} 

Many theories of quantum gravity suggest that 
a fundamental quantum description of spacetime 
induces violation of local Lorentz invariance 
in the effective low energy description of matter as a 
quantum field theory on a classical 
spacetime~\cite{AmelinoCamelia:2008qg}. 
Constraining local Lorentz violations observationally 
provides hence a potential constraint on quantum theories of gravity. 
Despite the large disparity 
between experimentally accessible energy scales and the Planck scale, 
which is usually thought to characterise quantum gravity phenomena, 
such constraints can be nontrivial: 
in quantum field theory, 
it is possible for the effective low energy theory 
to bear unexpected imprints of the theory's 
high energy structure~\cite{Collins:2004bp,Polchinski:2011za}. 

In this paper we consider a Lorentz-violating scalar field theory 
with the dispersion relation 
\begin{align}
\omega_{|\k|} = |\k| f\bigl(|\k|/\M\bigr)
\ , 
\label{eq:disprel-gen}
\end{align}
where we have set $c = \hbar = 1$, 
$\k$ is the spatial momentum, 
$\omega_{|\k|}$ is the energy, 
$\M$ is a positive constant of dimension energy,
and the dimensionless function $f$ satisfies $f(g) \to 1$ as $g\to0$. 
For $|\k|/\M \to 0$, the dispersion relation 
becomes that of a relativistic massless field, 
$\omega_{|\k|} = |\k|$, 
and we may hence think of $\M$ as the characteristic energy scale of 
Lorentz violation. 

Crucially, we assume that $f$ dips below unity somewhere. 
An example is the low energy sector of 
a scalar field quantised in the polymer
quantisation framework 
\cite{Ashtekar:2002sn,Ashtekar:2002vh,Husain:2010gb,Hossain:2010eb,Hossain:2009ru,Seahra:2012un}, 
motivated by loop quantum gravity~\cite{Rovelli-book,Thiemann-book}. 
Similar dips occur also in condensed matter systems, including 
the roton minimum of Helium~3~\cite{vollhardt-woelfle}. 

In four-dimensional Minkowski spacetime, 
an inertial Unruh-DeWitt detector 
\cite{Unruh:1976db,DeWitt:1979} coupled to a quantum field with these properties 
was shown in 
\cite{Kajuri:2015oza}
to be able to undergo spontaneous excitations, 
and it was shown in 
\cite{Husain:2015tna} that the detector experiences drastic 
Lorentz-violating excitations and de-excitations at 
arbitrarily low energies whenever the detector's speed in the preferred frame exceeds 
the critical value that equals the infimum of~$f$. 
The purpose of the present paper is to generalise these observations in two ways. 

First, we show that the inertial detector experiences a similarly 
drastic low energy Lorentz violation in Minkowski spacetimes of 
all dimensions greater than three, 
and a less drastic but still large low energy Lorentz violation in dimension three. 
In two dimensions, by contrast, the low energy Lorentz violation 
is suppressed by the factor~$1/\M$, 
except for a finetuned resonance that occurs 
when the speed is very close to the critical value.  

Second, we show analytically that a detector in uniform circular 
motion in three-dimensional Minkowski spacetime 
experiences a large low energy Lorentz violation above the same critical 
speed as in inertial motion, and we exhibit numerical results 
about the detailed form of these violations. 

We also perform the analytic groundwork for investigating 
circular motion in all dimensions above three, 
generalising the four-dimensional work in~\cite{Stargen:2017xii}. 
Generalising our three-dimensional circular motion analytic and numerical techniques 
to four dimensions and beyond would require an improved control of the generalised 
hypergeometric functions that appear, analytically in terms of uniform asymptotic expansions, 
and numerically in terms of accurate numerical evaluation 
in regimes where certain combinations of the parameters are large. 

While we do not have in mind a concrete experimental setup, 
we anticipate the results to have relevance 
not just for constraining fundamental quantum theories of gravity 
but also for designing analogue spacetime laboratory experiments, 
where the effective spacetime dimensionality of the 
system is often different from four 
\cite{Belgiorno:2010wn,Weinfurtner:2010nu,Steinhauer:2015saa,Torres:2016iee,Leonhardt:2017lwm}. 
For a selection of experimentally-motivated analyses of circular motion, see 
\cite{Bell:1982qr,Bell:1986ir,Leinaas:1998tu,Unruh:1998gq,Retzker-circular,Jin:2014coa,Jin:2014spa}. 

We begin in Section \ref{sec:setup} with a review 
of an Unruh-DeWitt detector coupled linearly to a 
scalar field with the modified dispersion relation~\eqref{eq:disprel-gen}, 
assuming the field to be in its Fock-like vacuum and the detector 
to be on a worldline on which the Fock-like vacuum appears stationary. We work within 
first-order perturbation theory, 
taking first the limit of weak interaction and then the limit of long interaction time. 
Section \ref{sec:inertial} addresses inertial motion, Section 
\ref{sec:helical2+1} addresses circular motion in $2+1$ dimensions, and Section 
\ref{sec:helical-n+1} addresses circular motion in dimensions 3+1 and greater. 
Section \ref{sec:conclusions}
provides brief concluding remarks. 
The proof of Theorem~\ref{theorem:2+1:limittheorem},
characterising the low energy limit of circular motion in $2+1$ dimensions, 
is deferred to the Appendix. 

We use units in which $c = \hbar = 1$. $\Theta$ denotes the Heaviside theta-function, 
\begin{align}
\Theta (x) 
= 
\begin{cases}
1&
\text{for $x>0\,$,}
\\
0 
&
\text{for $x\le0\,$.}
\end{cases}
\label{eq:heaviside-def}
\end{align}
$\lceil \cdot \rceil$ denotes the ceiling function, 
$\lceil x \rceil = $ the 
smallest integer greater than or equal to~$x$. 
$o(1)$~denotes a quantity that goes to 
zero in the limit under consideration.

\section{Field and detector\label{sec:setup}}

In this section we set up the notation for the field and 
the detector with which the field is probed. 

\subsection{Field\label{subsec:field}}

We work in $(n+1)$-dimensional Minkowski spacetime, where $n=1,2,\ldots$. 
We write the metric in the distinguished reference frame as
\begin{align}
ds^2 = -dt^2 + d\x^2 = - dt^2 + {(dx^1)}^2 + \cdots + {(dx^n)}^2
\ . 
\end{align}

We consider a scalar field $\phi$ 
with the dispersion relation~\eqref{eq:disprel-gen}, 
where $\k$ is the spatial momentum, 
$\omega_{|\k|}$ is the energy, 
$\M$ is a positive constant of dimension energy, 
and $f$ is a dimensionless function of a non-negative dimensionless variable. 
We assume that $f$ is smooth and positive-valued, and $f(g) \to 1$ as $g\to0$. 

We assume that $\phi$ admits a decomposition into spatial Fourier modes labelled by~$\k$, 
such that each mode with spatial momentum $\k \ne \mathbf{0}$ is a harmonic 
oscillator with the angular frequency $\omega_{|\k|}$~\eqref{eq:disprel-gen}, 
and the mode with $\k=\mathbf{0}$ is of measure zero and does not contribute to the decomposition. 
We include in the Fourier decomposition the density-of-states 
weight factor
\begin{align}
\rho_{|\k|} = 
\frac{d\bigl(|\k|/\M\bigr)}{\sqrt{{(2\pi)}^n |\k|}}
\ ,  
\label{eq:density-gen}
\end{align}
where $d$ is a dimensionless complex-valued function 
of a non-negative dimensionless variable. 
We assume that $d$ is smooth and nowhere vanishing, and 
$d(g) \to 1/\sqrt{2}$ as $g\to0$. 
In the limit $\M\to\infty$ with fixed~$\k$, we then have 
$\omega_{|\k|} \to |\k|$ and $\rho_{|\k|} \to \bigl(2{(2\pi)}^n |\k|\bigr)^{-1/2}$. 
The mode-by-mode $\M\to\infty$ limit of the field  
is hence the usual, Lorentz-invariant, massless scalar field. 

If $\phi$ is viewed strictly as a scalar field, 
$d$ and $f$ are related by $d(g) =1/\sqrt{2 f(g)}$. We shall keep $f$ and $d$ independent, 
as this will also maintain applicability to situations where the scalar 
field is an effective low energy limit of a more fundamental theory. 
One such situation is the low energy limit of the polymer quantised 
scalar field~\cite{Hossain:2010eb,Hossain:2009ru,Seahra:2012un}. 

Finally, we introduce the crucial assumption that $f$ dips below unity somewhere. 
For technical concreteness, we assume that the only stationary point of 
$f$ is a global minimum at $g = g_c>0$. 
Writing $f_c = f(g_c)$, we then have $0<f_c<1$. 
This is satisfied by the $f$ that emerges 
from the low energy limit of the polymer quantised scalar
field~\cite{Hossain:2010eb,Hossain:2009ru,Seahra:2012un}.

\subsection{Detector\label{subsec:detector}}

We probe the field by coupling it linearly to a spatially 
pointlike two-level quantum system 
known as the Unruh-DeWitt detector~\cite{Unruh:1976db,DeWitt:1979}. 
This coupling models an atom interacting with the 
electromagnetic field when angular momentum interchange 
is negligible~\cite{MartinMartinez:2012th,Alhambra:2013uja}, 
and it has been widely used to analyze motion effects in quantum field theory. 
We shall here just summarise the properties needed for the later sections. 
Textbook accounts can be found in 
\cite{Birrell:1982ix,Wald:1995yp}
and recent reviews in~\cite{Crispino:2007eb,Hu:2012jr,Martin-Martinez:2014gra}. 

We consider a detector that moves on the prescribed worldline 
\begin{align}
{\sf{x}}(\tau) = \bigl(t(\tau), \x(\tau)\bigr)
\ , 
\end{align}
parametrised by the proper time~$\tau$. 
We set the field initially in its Fock vacuum. 

In first-order
perturbation theory, the probability of the detector to make a
transition from the state with energy $0$ to the state with energy
$E$ (which may be positive, negative or zero) 
is proportional to the response function, 
\begin{align}
\mathcal{F}(E) = \int d\tau \, d\tau' \, 
\chi(\tau) \chi(\tau') \, 
e^{-i E (\tau-\tau')} 
\, \mathcal{W}(\tau,\tau')
\ ,
\label{eq:respfunction}
\end{align} 
where 
the switching function $\chi$ specifies how the interaction is turned on and off, 
$\mathcal{W}$ is the pullback of the field's Wightman
function to the detector's worldline, 
\begin{align}
\mathcal{W}(\tau,\tau') = G\bigl(t(\tau), \x(\tau); t(\tau'), \x(\tau') \bigr) 
\ ,
\label{eq:W-definition}
\end{align}
and the Wightman function is given by 
\begin{align}
G(t, \x; t', \x') 
= \int d^n\k \, | \rho_{|\k|}|^2 \, 
{e^{i {\k}\cdot(\x-\x') 
-i \omega_{|\k|} (t-t'- i \epsilon)}} 
\ , 
\label{eq:gen-propagator}
\end{align}
where the distributional character is encoded 
in the limit $\epsilon\to0_+$. 
The constant of proportionality is quadratic in the 
coupling constant and depends on the 
internal structure of the detector, 
but it is independent of $E$ and of the detector's trajectory, 
and we shall drop this constant from now on. 

We consider detector trajectories for which the Fock vacuum is stationary, 
in the sense that 
$\mathcal{W}(\tau,\tau')$ depends on its arguments
only through the difference $\tau-\tau'$. We may then 
convert $\mathcal{F}$ into  
the transition rate per unit time by passing 
to the limit of adiabatic switching and 
factoring out the effective total
duration of the detection. This procedure 
has significant conceptual and technical 
subtlety~\cite{Hu:2012jr,Satz:2006kb,Louko:2007mu,Fewster:2016ewy}, 
but the outcome is that the transition rate is 
proportional to 
\begin{align}
\mathcal{F}_{\rm rate}(E) = \int_{-\infty}^{\infty} ds 
\, 
e^{-i E s} \, 
\mathcal{W}(s,0) 
\ . 
\label{eq:transrate-nosigma}
\end{align}
In what follows we shall work with~\eqref{eq:transrate-nosigma}, 
dropping the ``rate'' subscript.

\section{Inertial motion\label{sec:inertial}}

In this section we consider a detector on the inertial worldline 
\begin{align}
{\sf{x}}(\tau) =
\bigl(t(\tau), \x(\tau) \bigr) 
= \bigl(\tau \cosh\beta, 0,\ldots,0,\tau\sinh\beta \bigr)
\ , 
\label{eq:inertial-trajectory}
\end{align}
where $\beta\ge0$ is the rapidity with respect to the distinguished 
inertial frame. For presentational simplicity we assume 
$\beta>0$, but it can be verified that there is no discontinuity as $\beta\to0$.

\subsection{Spacetime dimension $2+1$ and greater}

We start in spacetime dimensions $2+1$ and greater, so that $n\ge2$. 
The case $n=1$ will be addressed in subsection~\ref{subsec:inertial1+1}. 

Inserting the trajectory 
\eqref{eq:inertial-trajectory}
into \eqref{eq:W-definition}, 
\eqref{eq:gen-propagator}, 
and~\eqref{eq:transrate-nosigma}, 
we obtain 
\begin{align}
\mathcal{F}(E) &= 
\int_{-\infty}^{\infty} ds 
\int d^n\k \, 
|\rho_{|\k|}|^2 \, 
e^{-i(E + \omega_{|\k|} \cosh\beta- k_n \sinh\beta)s}
\notag
\\
&= 
\Omega_{n-2}
\int_{-\infty}^{\infty} ds 
\int_0^\infty dK \, K^{n-1} \, 
|\rho_K|^2 
\int_0^\pi d\theta \, {(\sin\theta)}^{n-2} \, 
e^{-i(E + \omega_K \cosh\beta- K \sinh\beta \cos\theta)s}
\notag
\\
&= 
2\pi \Omega_{n-2}
\int_0^\infty dK \, K^{n-1} \, 
|\rho_K|^2 
\int_0^\pi d\theta \, {(\sin\theta)}^{n-2} \, 
\delta(E + \omega_K \cosh\beta- K \sinh\beta \cos\theta)
\notag
\\
&= 
\frac{2\pi \Omega_{n-2}}{\sinh\beta} 
\int_0^\infty dK \, K^{n-2} \, 
|\rho_K|^2 
{\left(1 - \frac{{(E + \omega_K \cosh\beta)}^2}{K^2 \sinh^2\!\beta}\right)}^{(n-3)/2}
\notag
\\
& \hspace{15ex}
\times \Theta \! \left(1 - \frac{|E + \omega_K \cosh\beta)|}{K \sinh\beta}\right)
\notag
\\
&= 
\frac{2 \M^{n-2}}{{{(4\pi)}^{(n-1)/2} \, \Gamma \! \left(\frac{n-1}{2}\right) (\sinh\beta)}^{n-2}}
\notag
\\
& \hspace{3ex}
\times 
\int_0^\infty dg \, 
|d(g)|^2 
\, 
\Theta \bigl(g\sinh\beta - |(E/\M) + g f(g) \cosh\beta | \bigr) 
\notag
\\
& \hspace{9ex}
\times 
\left(g^2\sinh^2\!\beta - {\bigl[(E/\M) + g f(g) \cosh\beta\bigr]}^2 \right)^{(n-3)/2} 
\ . 
\label{eq:Finert-gendim}
\end{align}
The second equality follows by writing $\k$ in spherical coordinates, 
such that $K = |\k|$ and $k_n = K\cos\theta$ with $\theta\in[0,\pi]$, 
and noting that the integral over the remaining $d-2$ angles brings out the area 
of the unit sphere in Euclidean~$\BbbR^{n-1}$, 
\begin{align}
\Omega_{n-2} = \frac{2 \pi^{(n-1)/2}}{\Gamma \! \left(\frac{n-1}{2}\right)}
\ . 
\label{eq:spherearea-formula}
\end{align}
The next two equalities follow by performing the integrals over $s$ and~$\theta$, respectively. 
In the final equality we have used \eqref{eq:spherearea-formula} and 
introduced the new integration variable $g = K/\M$. 

Recall \cite{Birrell:1982ix}
that the transition rate of the Lorentz-invariant massless scalar 
field can be obtained from the first line of 
\eqref{eq:Finert-gendim}
by setting $f(g)=1$ and $d(g) = 1/\sqrt{2}$, with the result 
\begin{align}
\Finnormal(E) = 
\frac{{(-E)}^{n-2} \Theta(-E)}{2 \, {{(4\pi)}^{(n-2)/2}} \, \Gamma \! \left(\frac{n}{2}\right)}
\ , 
\label{eq:Finert-gen-inv}
\end{align}
where the subscript $0$ indicates the unmodified massless scalar 
field and the superscript $in$ indicates inertial motion. 
Note that $\Finnormal(E)$ vanishes for $E>0$: 
the Lorentz-invariant 
field induces no spontaneous excitations in the detector. 

We wish to compare $\mathcal{F}(E)$ 
to $\Finnormal(E)$ in the low energy limit, 
$|E|/\M \ll 1$. 

Suppose first that $n\ge3$. The case $n=3$ was considered in~\cite{Husain:2015tna}, 
and inspection of \eqref{eq:Finert-gendim} shows that the techniques used therein 
adapt readily to all $n\ge3$. 
When $\beta < \beta_c = \arctanh f_c$, $\mathcal{F}(E)$ vanishes for 
$E>0$, and the corrections to $\Finnormal(E)$ are small for $E<0$ with $|E|/\M \ll 1$. 
However, when $\beta > \beta_c$, 
$\mathcal{F}(E)$ is of the order of~$\M^{n-2}$ whenever $|E|/\M \ll 1$, 
both for $E>0$ and for $E<0$. 
This is a drastic low energy Lorentz violation, 
as noted for $n=3$ in~\cite{Husain:2015tna}. 

Suppose then that $n=2$, in which case 
\eqref{eq:Finert-gendim} reduces to 
\begin{align}
\mathcal{F}(E) 
&= 
\frac{1}{\pi}
\int_0^\infty dg \, 
|d(g)|^2 
\frac{\Theta\bigl(g\sinh\beta - |(E/\M) + g f(g) \cosh\beta | \bigr)}
{\sqrt{g^2\sinh^2\!\beta - {\bigl[(E/\M) + g f(g) \cosh\beta\bigr]}^2}}
\ . 
\label{eq:Finert-2+1}
\end{align}
When $\beta < \beta_c$, $\mathcal{F}(E)$ again vanishes for 
$E>0$, and the corrections to $\Finnormal(E)$ 
are small for $E<0$ with $|E|/\M \ll 1$. 
When $\beta > \beta_c$, 
$\mathcal{F}(E)$ is nonvanishing and of order unity whenever $|E|/\M \ll 1$, 
and for $E<0$ it is larger than $\Finnormal(E)$ by a factor of order unity. 
This amounts to a low energy Lorentz violation that is large, 
in the sense of having a magnitude 
that is comparable to the Lorentz-invariant de-excitation rate. 
However, the violation does not involve enhancement 
factors by positive powers of $\M/|E|$ as in $n\ge3$. 

\begin{figure}[p]
\centering
\includegraphics[width=0.8\textwidth]{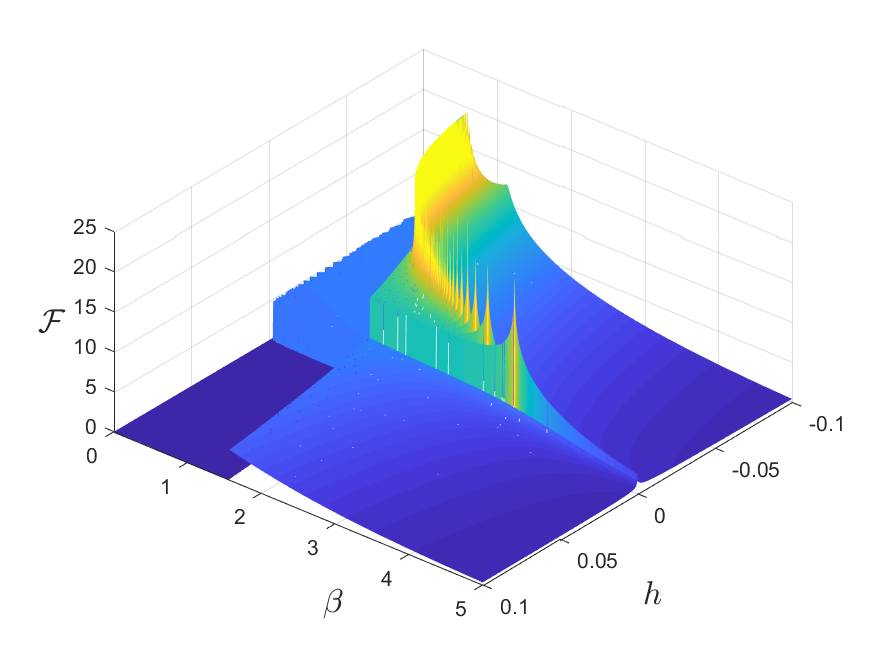}
\caption{A perspective plot of the $n=2$ transition rate $\mathcal{F}(E)$ \eqref{eq:Finert-2+1}, 
as a function of $\beta$ and $h = E/\M$, for the dispersion relation and 
density of states~\eqref{eq:mock-up-fullset}. 
In the low energy regime, $|E|/\M \ll 1$, 
large deviations from the Lorentz-invariant transition rate \eqref{eq:Finert-gen-inv}
are apparent when $\beta$ increases above the critical value $\beta_c \approx 1.3675$.}
\label{fig:2d-inertial-surface}
\end{figure}

\begin{figure}[p]
\centering
\includegraphics[width=0.7\textwidth]{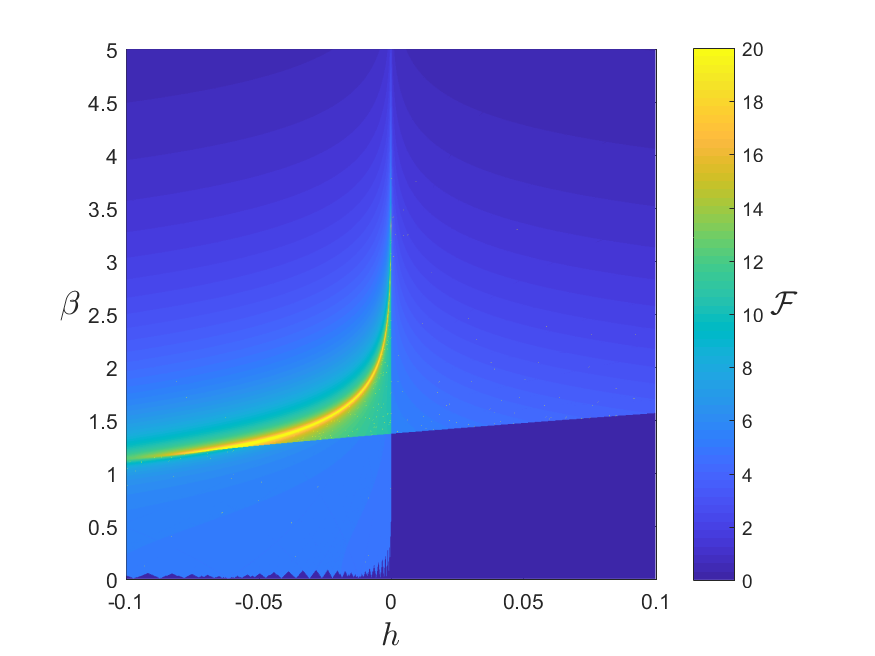}
\caption{As in Figure \ref{fig:2d-inertial-surface} but in a topdown colour (grayscale) plot.}
\label{fig:2d-inertial-topdown}
\end{figure}

\subsection{Graphics for spacetime dimension $2+1$}

Figures \ref{fig:2d-inertial-surface} and \ref{fig:2d-inertial-topdown}
give plots of the $n=2$ transition rate 
$\mathcal{F}(E)$ \eqref{eq:Finert-2+1}, 
as a function of $\beta$ and~$E/\M$. 
The plots use the dispersion relation and the density-of-states given by 
\begin{subequations}
\label{eq:mock-up-fullset}
\begin{align}
f(g) &= \sqrt{(1 - f_c^2) (g-1)^2 + f_c^2}
\ , 
\label{eq:mock-up-dispersion}
\\
f_c &= 0.8781
\ , 
\label{eq:fc-valuechoice}
\\
d(g) &= 1/\sqrt{2 f(g)}
\ . 
\end{align}
\end{subequations}
We chose this dispersion relation for its numerical 
amenability and its qualitative 
similarity to the dispersion relation that emerges 
from the low energy limit of the polymer quantised scalar
field~\cite{Hossain:2010eb,Hossain:2009ru,Seahra:2012un}. 

When $|E|/\M \ll 1$, the jump as $\beta$ increases above 
$\beta_c \approx 1.3675$ is plain in the plots. 

The large spike in the plots 
at negative $E/\M$ 
is a genuine infinity that occurs 
along a curve in the $(\beta, E/\M)$ plane. 
(That the perspective plot in Figure 
\ref{fig:2d-inertial-surface} shows individual spikes, rather than a curve, 
is a numerical artefact.) 
The mathematical reason for this de-excitation resonance is that for values 
of $(\beta, E/\M)$ on this curve, 
the function of 
$g$ under the square root in \eqref{eq:Finert-2+1} has a quadratic minimum 
at value zero. 
The spike starts to occur for $\beta \gtrsim 1.2059$, and it exists for 
$-0.06917 \lesssim E/\M < 0$. This de-excitation resonance has no counterpart for $n\ge3$ 
because the exponent $(n-3)/2$ in  
\eqref{eq:Finert-gendim} is nonnegative for $n\ge3$.

\subsection{Spacetime dimension $1+1$\label{subsec:inertial1+1}}

Finally, we consider spacetime dimension $1+1$, so that $n=1$. 

The Wightman function \eqref{eq:gen-propagator} 
is now infrared divergent. However, ignoring this divergence
and proceeding formally through steps similar to those in~\eqref{eq:Finert-gendim}, 
we obtain 
\begin{align}
\mathcal{F}(E) &= 
\frac{1}{\M} \int_0^\infty \frac{dg}{g} \, |d(g)|^2 
\Bigl[
\delta\bigl( (E/\M) + g f(g) \cosh\beta - g\sinh\beta \bigr)
\notag
\\
& \hspace{23ex}
+ \delta\bigl( (E/\M) + g f(g) \cosh\beta + g\sinh\beta \bigr)
\Bigr]
\ ,   
\label{eq:Finert-1+1-intform}
\end{align}
from which the infrared divergence has disappeared at the 
step of interchanging the integrals. 
Performing the integral in \eqref{eq:Finert-1+1-intform} gives
\begin{align}
\mathcal{F}(E) &= \sum_j \frac{{|d(g_j)|}^2}{\bigl| E - \M g_j^2 f'(g_j)\bigr|}
\ , 
\label{eq:Finert-1+1-sumform}
\end{align}
where $g_j$ are the solutions to  
\begin{align}
0 = \frac{E}{\M} + g f(g) \cosh\beta \mp g \sinh\beta
\ ,  
\label{eq:1+1-gj-eq}
\end{align}
including both signs. 
The upper (lower) sign in \eqref{eq:1+1-gj-eq} comes from 
the first (second) Dirac delta in~\eqref{eq:Finert-1+1-intform}. 
We adopt \eqref{eq:Finert-1+1-intform}, 
or equivalently \eqref{eq:Finert-1+1-sumform} and~\eqref{eq:1+1-gj-eq}, as the definition 
of~$\mathcal{F}(E)$. The same answer may be obtained by introducing both an infrared cutoff 
and a long time cutoff and taking a suitable limit, 
via a procedure discussed in the context of the Unruh effect in~\cite{Takagi:1986kn}. 

The transition rate of the Lorentz-invariant massless scalar field is obtained by setting 
$f(g)=1$ and $d(g) = 1/\sqrt{2}$ in 
\eqref{eq:Finert-1+1-sumform} and~\eqref{eq:1+1-gj-eq}, with the outcome 
\begin{align}
\Finnormal(E) = 
\frac{\Theta(-E)}{(-E)}
\ . 
\label{eq:Finert-1+1-inv}
\end{align}
Note that \eqref{eq:Finert-1+1-inv} fits in the $n\ge2$ pattern of~\eqref{eq:Finert-gen-inv}, 
in the sense that setting 
$n=1$ in \eqref{eq:Finert-gen-inv} agrees with~\eqref{eq:Finert-1+1-inv}. 

We wish to compare $\mathcal{F}(E)$ with 
$\Finnormal(E)$ in the low energy limit, 
$|E|/\M \ll 1$. 

\begin{figure}[t]
\centering
\includegraphics[width=0.6\textwidth]{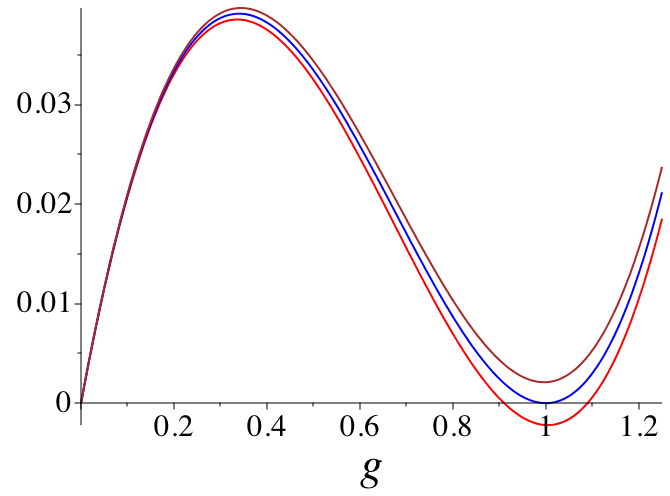}
\caption{Plot of the function $g f(g) \cosh\beta - g \sinh\beta$, 
with $f$ given in~\eqref{eq:mock-up-fullset}. The middle curve is for 
$\beta = \beta_c \approx 1.3675$, the top curve is for 
$\beta = 1.363 < \beta_c$, and the bottom curve is for $\beta = 1.372 > \beta_c$.}
\label{fig:1plus1curves}
\end{figure}

Suppose first that $\beta < \beta_c$. 
For $E>0$, 
$\mathcal{F}(E)$ vanishes. For 
$E<0$, $\mathcal{F}(E)$ tends to $\Finnormal(E)$ 
for sufficiently small $|E|/\M$, but the sense of ``sufficiently small'' 
depends on~$\beta$: 
if $\beta$ is close to~$\beta_c$, 
$\mathcal{F}(E)$ has a divergent peak at the negative 
value of $E$ for which the argument of the first 
delta-function in \eqref{eq:Finert-1+1-intform} has the stationary value zero, 
and this divergent peak can be made to occur at arbitrarily low 
$|E|/\M$ by taking $\beta$ close to~$\beta_c$. 
We illustrate this phenomenon in 
Figure~\ref{fig:1plus1curves}, where the top curve shows the 
function $g f(g) \cosh\beta - g \sinh\beta$ with the 
$f$ given in~\eqref{eq:mock-up-fullset} and $\beta$ slightly below~$\beta_c$. 

Suppose then that $\beta > \beta_c$. 
The first delta-function in \eqref{eq:Finert-1+1-intform} 
brings in additional contributions to $\mathcal{F}(E)$ whenever 
$|E|/\M \ll 1$, regardless the sign of~$E$. 
These contributions come from $g$ of order unity, 
and for sufficiently small $|E|/\M$ they are of
order~$1/\M$, and hence small compared with $\Finnormal(E)$, 
but the sense of ``sufficiently small'' again depends on~$\beta$: 
if $\beta$ is close to~$\beta_c$, 
$\mathcal{F}(E)$ now has a divergent peak 
at the positive value of $E$ for which the argument of the first 
delta-function in \eqref{eq:Finert-1+1-intform} 
has the stationary value zero, and this divergent 
peak can be made to occur at arbitrarily low 
$|E|/\M$ by taking $\beta$ close to~$\beta_c$. 
This phenomenon is illustrated by the bottom curve in 
Figure~\ref{fig:1plus1curves}. 

Finally, suppose that $\beta$ is strictly equal to~$\beta_c$, 
as illustrated in the middle curve in Figure~\ref{fig:1plus1curves}. 
For $E>0$, $\mathcal{F}(E)$ vanishes. For $E<0$ and $|E|/\M \ll 1$, 
the first delta-function in \eqref{eq:Finert-1+1-intform} 
brings in additional contributions from $g$ close to~$g_c$, 
and using 
\eqref{eq:Finert-1+1-sumform} and~\eqref{eq:1+1-gj-eq} we find that these contributions 
are of order ${(-E \M)}^{-1/2}$, which is small compared 
with $\Finnormal(E)$ by the factor $\sqrt{-E/\M}$. 

To summarise, these observations show that the low energy Lorentz violation 
is small, except for a finetuned de-excitation (excitation) 
peak when $\beta$ is finetuned to be narrowly below (above)~$\beta_c$. 

As a final remark, we note that if $g f'(g) + f(g)$ is not strictly positive and 
$\beta$ is sufficiently small, 
there are also divergent de-excitation peaks that come from the stationary zeroes of the 
argument of the second delta-function in~\eqref{eq:Finert-1+1-intform}. 
However, these peaks do not occur at low energies.

\section{Circular motion in $2+1$ dimensions\label{sec:helical2+1}}

In this section we consider a detector in uniform circular motion in spacetime dimension $2+1$. 
The worldline is 
\begin{align}
{\sf{x}}(\tau) =
\bigl(t(\tau), \x(\tau) \bigr) 
= \bigl(\gamma \tau, R \cos(\gamma\Omega\tau),  R \sin(\gamma\Omega\tau)\bigr)
\ , 
\label{eq:helical-trajectory2+1}
\end{align}
where $R$ and $\Omega$ are positive parameters satisfying $R\Omega < 1$, 
and $\gamma = 1/\sqrt{1 - R^2 \Omega^2}$.  
$R$~is the radius of the orbit, and $\Omega$ is the angular velocity 
in the preferred Lorentz frame. 
The worldline is an orbit of the Killing vector 
$\partial_t + \Omega (x^1 \partial_{2} - x^2 \partial_{1})$,  
and its scalar proper acceleration is~$R\Omega^2\gamma^2$.  

Inserting the trajectory 
\eqref{eq:helical-trajectory2+1}
into \eqref{eq:W-definition} and~\eqref{eq:gen-propagator}, 
we obtain 
\begin{align}
\mathcal{W}(s,0) 
&= 
\int d^2\k \, 
|\rho_{|\k|}|^2 \, 
e^{-i\omega_{|\k|} \gamma s + i R \{k_1 [\cos(\gamma\Omega s) -1] + k_2\sin(\gamma\Omega s)\}}
\notag
\\
&= 
\int_0^\infty dK \, K |\rho_{K}|^2 \, 
e^{-i \omega_{K} \gamma s}
\int_0^{2\pi}d\varphi \, e^{2i R K \sin(\gamma\Omega s/2) \sin(\varphi - \gamma\Omega s/2)}
\notag
\\
&= 
2\pi \int_0^\infty dK \, K |\rho_{K}|^2 \, 
e^{-i \omega_{K} \gamma s}
\, J_0 \bigl(2 R K \sin(\gamma\Omega s/2) \bigr)
\notag
\\
&= 
2\pi \int_0^\infty dK \, K |\rho_{K}|^2 \, 
e^{-i \omega_{K} \gamma s}
\sum_{m\in\BbbZ} J_m^2 (R K) \, e^{im\gamma\Omega s}
\ . 
\label{eq:W-2+1helical-initial}
\end{align}
At the second equality we have written 
$\k = (K\cos\varphi, K\sin\varphi)$, 
where $0\le K < \infty$ and $0 \le \varphi < 2\pi$. The third equality uses
10.9.1 in~\cite{dlmf}, 
and the fourth equality uses the identity 
\begin{align}
J_0(2a \sin x) = \sum_{m\in\BbbZ} J_m^2(a) \, e^{2imx}
\ , 
\end{align}
which can be verified using 6.681.6 in~\cite{grad-ryzh}. 

Substituting \eqref{eq:W-2+1helical-initial}
into~\eqref{eq:transrate-nosigma}, 
performing the integral over~$s$, using 
\eqref{eq:disprel-gen}
and~\eqref{eq:density-gen}, 
and writing $g = K/\M$, we find 
\begin{align}
\mathcal{F}(E) 
&= 
\frac{1}{\gamma}
\sum_{m\in\BbbZ}
\int_0^\infty dg \, |d(g)|^2 
\, 
J_m^2(\Mtilde g) \, 
\delta \! \left(g f(g) - \frac{1}{\Mtilde} 
\left(mv - \frac{\Etilde}{\gamma}\right)\right)
\ , 
\label{eq:Fhelical2+1-final}
\end{align}
where $\Mtilde = R \M$, $\Etilde = RE$ and $v = R\Omega$. 
$v$ is the detector's speed in the preferred Lorentz frame, 
and the dimensionless parameters $\Mtilde$ and $\Etilde$ 
are respectively $\M$ and $E$ expressed in units of~$1/R$. 
Note that $0<v<1$. 

Recall that by assumption $f$ and $d$ are smooth, 
$f$ is strictly positive and 
$d$ is nowhere vanishing, and
$f(g) \to 1$ and $d(g) \to 1/\sqrt{2}$ as $g\to0$. 
Recall also that by assumption the only 
stationary point of $f$ is a global minimum at 
$g = g_c>0$, and we write $f_c = f(g_c)$, where $0<f_c<1$. 
We then have $f'(g) < 0$ for $0<g<g_c$ and $f'(g) > 0$ for $g>g_c$. 

To proceed, we make two additional assumptions. First, we assume 
that $gf'(g) + f(g)>0$ for $g>0$.  
It follows that $g f(g)$ is a strictly increasing function of~$g$. 
Performing the integral in~\eqref{eq:Fhelical2+1-final}, we then find 
\begin{align}
\mathcal{F}(E) 
&= 
\frac{1}{\gamma}
\sum_{m = \lceil \Etilde/(v\gamma)\rceil}^\infty 
\frac{{|d(g_m)|}^2}{g_m f'(g_m) + f(g_m)} 
\, 
J_m^2(\Mtilde g_m) 
\ , 
\label{eq:Fhelical2+1-sumform}
\end{align}
where $g_m$ is the unique solution to 
\begin{align}
g f(g) = \frac{1}{\Mtilde} \left(m v - \frac{\Etilde}{\gamma}\right)
\ . 
\label{eq:gm-def}
\end{align}
Second, we assume that $f(g)>1$ for sufficiently large~$g$, 
and that $|d(g)|$ is bounded. 

We wish to examine $\mathcal{F}(E)$ in the limit $\Mtilde\to\infty$. 
We find that the limit is qualitatively different for $0<v< f_c$ 
and $f_c < v < 1$, as given by the following theorem. 

\begin{thm}
\label{theorem:2+1:limittheorem}
Under the assumptions stated above: 
\begin{enumerate}
\item[(i)]
For $0<v<f_c$, $\mathcal{F}(E) \to \Fcircnormal(E)$ as $\Mtilde\to\infty$, where 
\begin{align}
\Fcircnormal(E) 
&= 
\frac{1}{2\gamma}
\sum_{m = \lceil \Etilde/(v\gamma)\rceil}^\infty 
J_m^2\bigl(mv - (\Etilde/\gamma)\bigr) 
\ . 
\label{eq:Fhelical2+1-normalfield}
\end{align}
\item[(ii)]
For $f_c < v < 1$, $\mathcal{F}(E) \to \Fcircnormal(E) + \Delta \mathcal{F}$ 
as $\Mtilde\to\infty$, where 
\begin{align}
\Delta \mathcal{F}
&= 
\frac{1}{\pi\gamma}
\int_{g_-}^{g_+}
dg 
\, \frac{{|d(g)|}^2}{g \, \sqrt{v^2 - 
f^2(g)}} 
\ ,  
\label{eq:DeltaFhelical2+1}
\end{align}
and $g_- \in (0,g_c)$ and $g_+ \in (g_c,\infty)$ are the 
unique solutions to 
$f(g) = v$ in the respective intervals. 
\end{enumerate}
\end{thm}

The proof of Theorem 
\ref{theorem:2+1:limittheorem}
is given in the Appendix. 

Part (i) of Theorem 
\ref{theorem:2+1:limittheorem}
says that the detector sees no low energy Lorentz violation when $v < f_c$: 
$\Fcircnormal(E)$ is the response for the usual massless scalar field, 
as is seen by comparing \eqref{eq:Fhelical2+1-sumform}
and~\eqref{eq:Fhelical2+1-normalfield}. 
The subscript $0$ in $\Fcircnormal$ indicates the usual massless scalar field and the 
superscript $circ$ indicates the circular trajectory. 

Part (ii) of Theorem 
\ref{theorem:2+1:limittheorem}
says that the detector sees a low energy Lorentz violation when $v > f_c$: 
this violation shows up as an increase in the excitation and de-excitation rates. 
The magnitude of the violation is similar to that which occurs in inertial motion 
in $2+1$ dimensions, found in Section~\ref{sec:inertial}. 

\begin{figure}[p]
\centering
\begin{tabular}{cc}
\hspace*{-2ex}\includegraphics[width=0.5\textwidth]{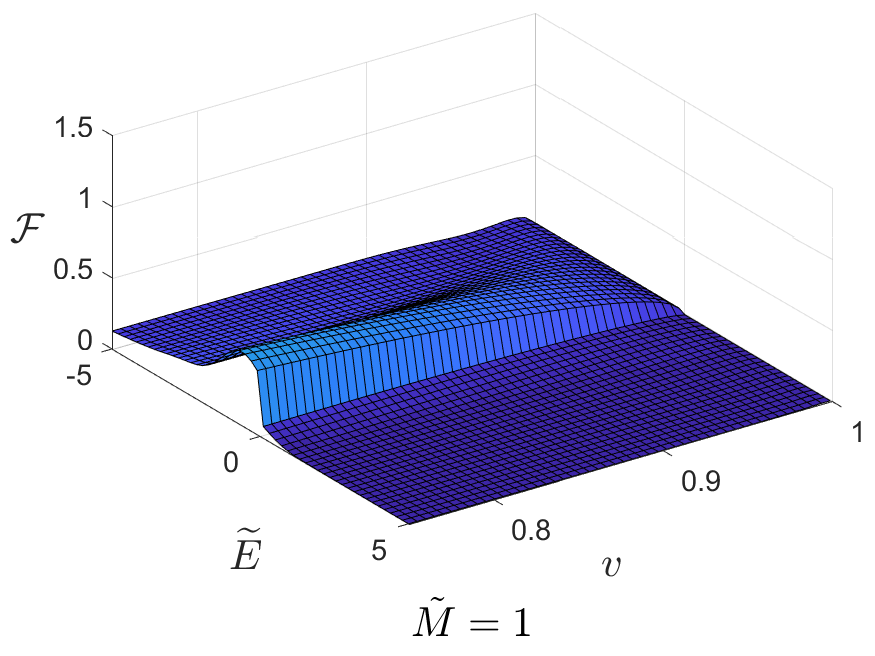}&%
\includegraphics[width=0.5\textwidth]{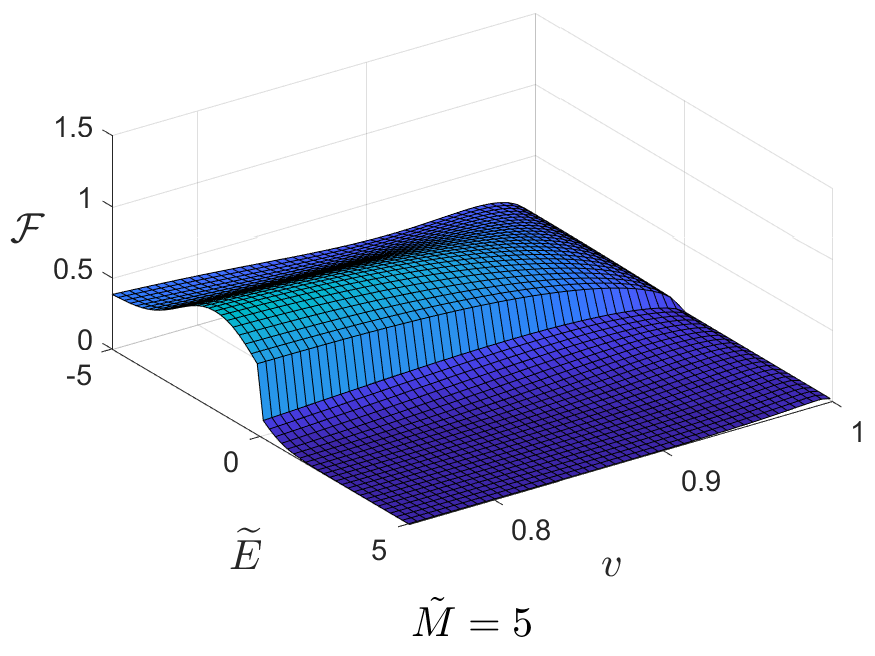}\\[2ex]
\hspace*{-2ex}\includegraphics[width=0.5\textwidth]{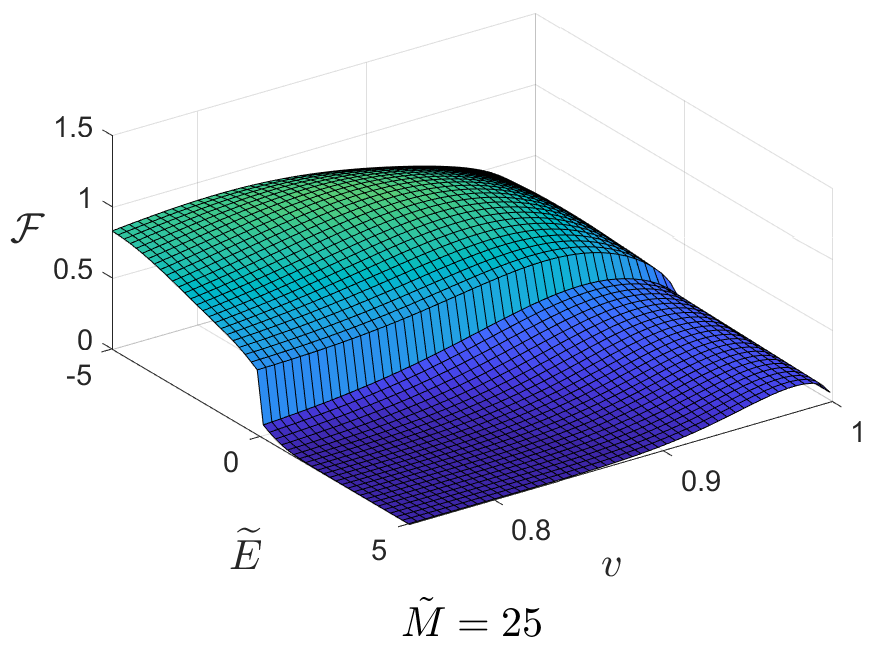}&%
\includegraphics[width=0.5\textwidth]{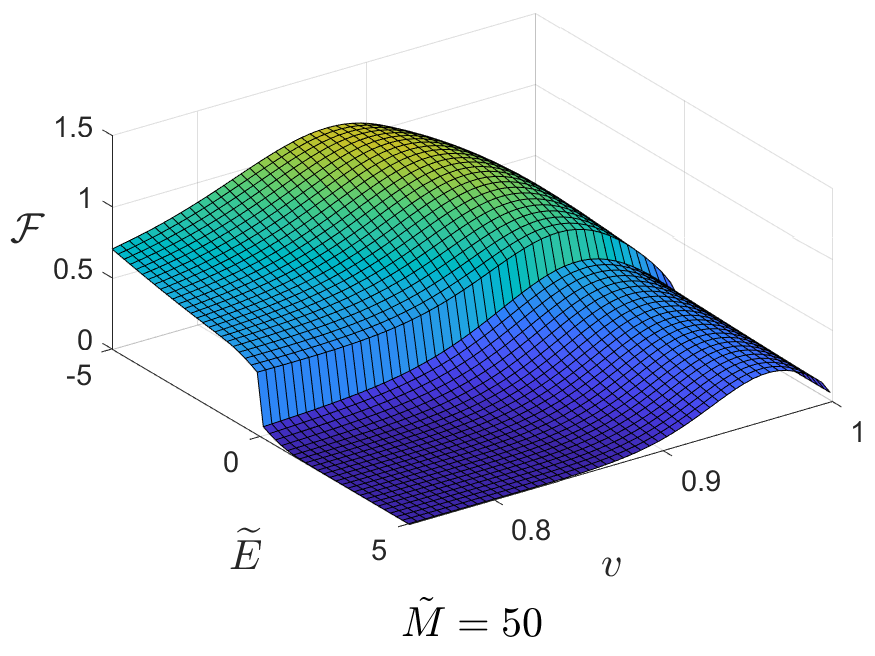}\\[2ex]
\hspace*{-2ex}\includegraphics[width=0.5\textwidth]{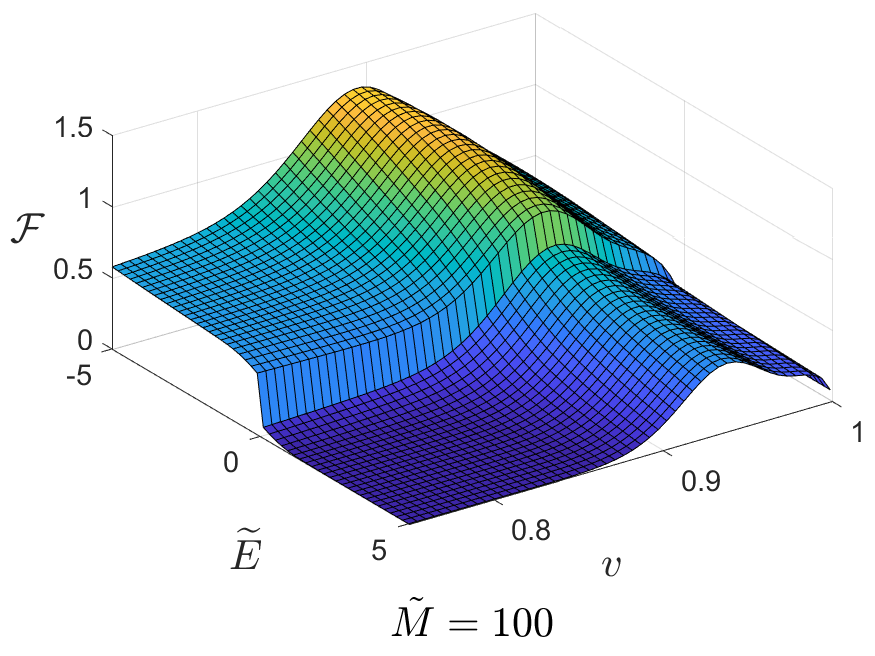}&%
\includegraphics[width=0.5\textwidth]{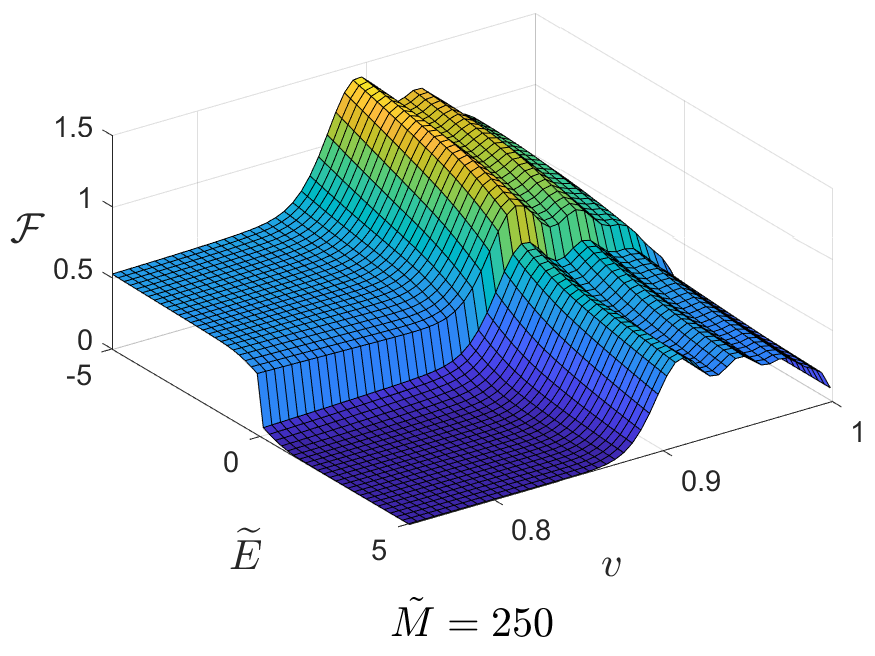}
\end{tabular}
\caption{Plots of $\mathcal{F}(E)$ \eqref{eq:Fhelical2+1-sumform} 
as a function of $\Etilde = RE$ and~$v$, for $f$ and $d$ given by~\eqref{eq:mock-up-fullset}, 
with $\Mtilde$ increasing from $1$ to $250$ as shown.}
\label{fig:2plus1helicalplotspart1}
\end{figure}

\begin{figure}[p]
\centering
\begin{tabular}{cc}
\hspace*{-2ex}\includegraphics[width=0.5\textwidth]{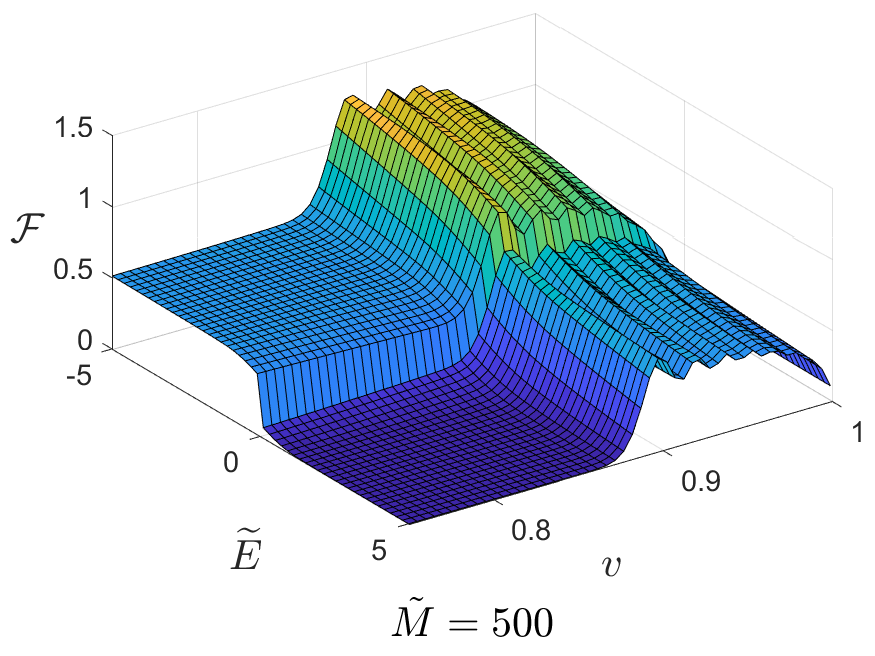}&%
\includegraphics[width=0.5\textwidth]{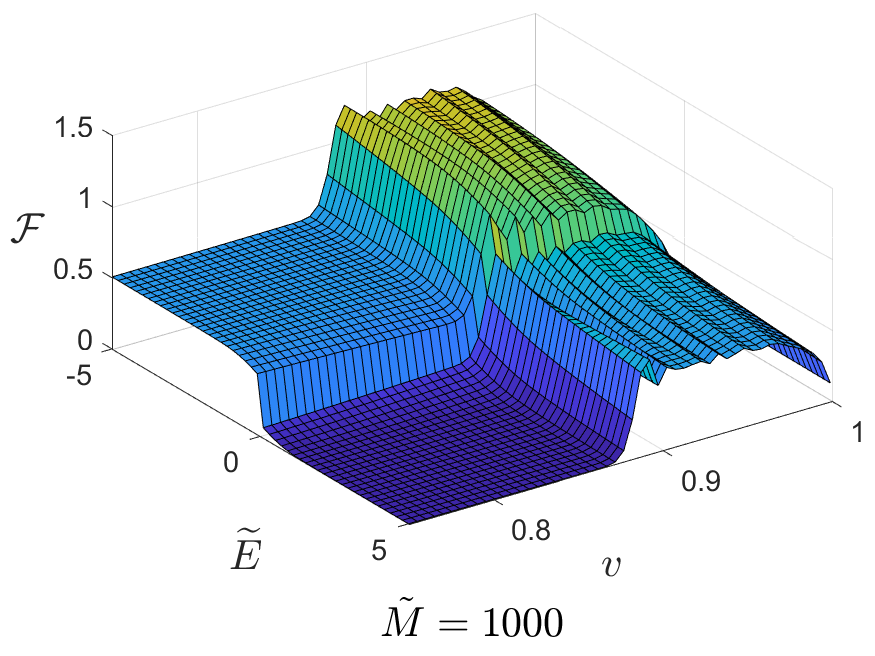}\\[2ex]
\hspace*{-2ex}\includegraphics[width=0.5\textwidth]{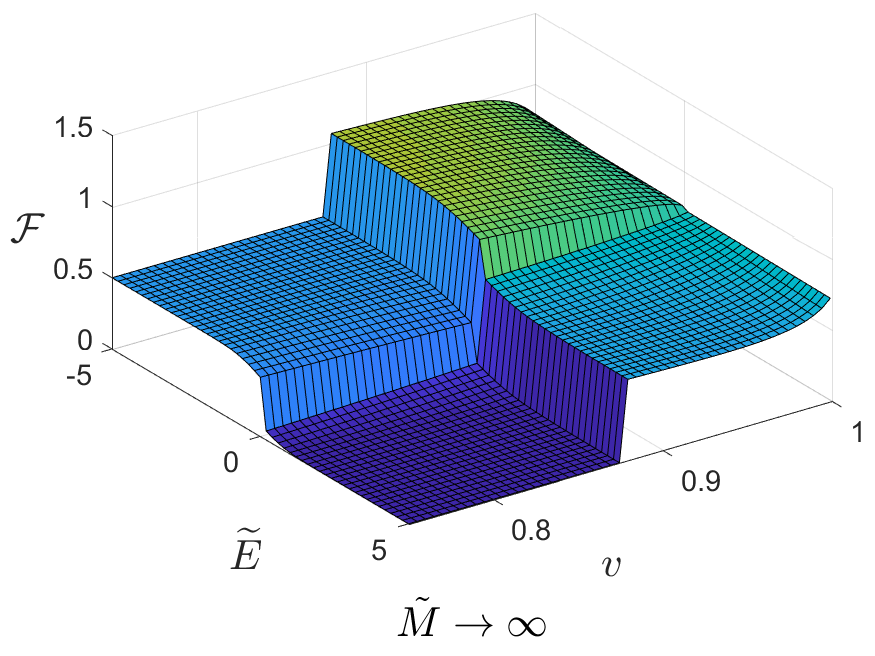}&%
\includegraphics[width=0.5\textwidth]{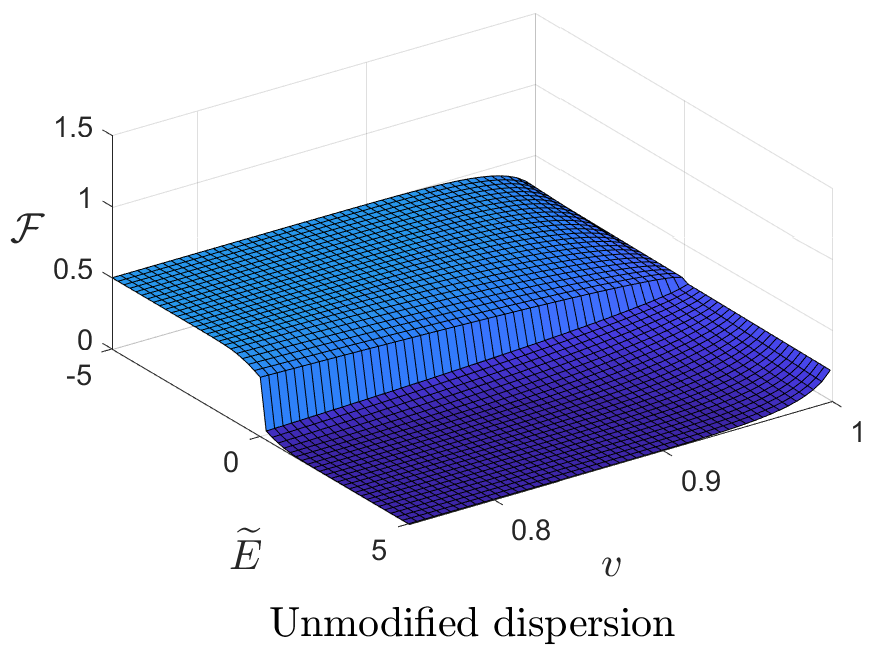}
\end{tabular}
\caption{The first two plots continue from Figure~\ref{fig:2plus1helicalplotspart1}, 
with $\Mtilde$ increasing to $1000$. The third plot shows the $\Mtilde\to\infty$ limit, 
given by Theorem~\ref{theorem:2+1:limittheorem}. 
The last plot shows the transition rate for the ordinary 
massless scalar field, 
given by $\Fcircnormal(E)$~\eqref{eq:Fhelical2+1-normalfield}. 
For large~$\Mtilde$, 
a large low energy deviation from the ordinary massless 
scalar field transition rate is clear from the plots when 
$v$ increases above the critical value 
$f_c = 0.8781$.}
\label{fig:2plus1helicalplotspart2}
\end{figure}

In Figures \ref{fig:2plus1helicalplotspart1} 
and \ref{fig:2plus1helicalplotspart2}
we give plots of~$\mathcal{F}(E)$, with  
the dispersion relation and the density of states given in~\eqref{eq:mock-up-fullset}, 
increasing $\Mtilde$ from $1$ to $1000$. [We note that \eqref{eq:mock-up-fullset} 
satisfies the technical assumptions of Theorem~\ref{theorem:2+1:limittheorem}.] 
The plots show how $\mathcal{F}(E)$ converges to the 
$\Mtilde\to\infty$ limit, given by Theorem \ref{theorem:2+1:limittheorem} 
and displayed in the penultimate plot 
in Figure~\ref{fig:2plus1helicalplotspart2}.  
Comparison with the unmodified dispersion relation transition rate, 
given in the last plot in Figure~\ref{fig:2plus1helicalplotspart2}, 
shows plainly the low energy Lorentz violation
when $v$ increases above~$f_c$. 

Two comments on the numerics are in order. First, 
when $v>f_c$ and $\Mtilde$ is large, 
the Lorentz-breaking contribution to the 
sum in \eqref{eq:Fhelical2+1-sumform}
comes from values of $m$ 
that are comparable to~$\Mtilde$, 
as seen from the proof of 
Part (ii) of 
Theorem \ref{theorem:2+1:limittheorem} in the Appendix. 
In the numerical evaluation of 
$\mathcal{F}(E)$ from \eqref{eq:Fhelical2+1-sumform}, 
we chose a summation cutoff that is high enough 
to include these values of~$m$. 
Second, we implemented the sum in \textsc{Matlab}, 
and we checked that the \textsc{Matlab} routine
for evaluating the Bessel functions in this range of $m$ is in agreement 
with the analytically known asymptotics, given by \eqref{eq:nuJnuz-expansion} 
in the Appendix. 

In conclusion, a detector in circular motion in the preferred inertial frame 
sees a large low energy Lorentz violation 
when its orbital speed 
exceeds the critical value~$f_c$. 
This low energy Lorentz violation is quite similar to the one seen by 
the inertial detector.

\section{Circular motion in spacetime dimensions $3+1$ and greater\label{sec:helical-n+1}}

In this section we consider a detector in uniform 
circular motion in spacetime dimensions $3+1$ and greater. 

The circular trajectory 
\eqref{eq:helical-trajectory2+1} generalises to 
\begin{align}
{\sf{x}}(\tau) =
\bigl(t(\tau), \x(\tau) \bigr) 
= \bigl(\gamma \tau, R \cos(\gamma\Omega\tau),  R \sin(\gamma\Omega\tau), 0, 0, \ldots\bigr)
\ . 
\label{eq:helical-trajectory-high}
\end{align}
Inserting \eqref{eq:helical-trajectory-high}
into \eqref{eq:W-definition} and~\eqref{eq:gen-propagator}, 
where now $n\ge3$, 
we find that \eqref{eq:W-2+1helical-initial} is replaced by 
\begin{align}
\mathcal{W}(s,0) 
&= 
\int d^n\k \, 
|\rho_{|\k|}|^2 \, 
e^{-i\omega_{|\k|} \gamma s + i R \{k_1 [\cos(\gamma\Omega s) -1] + k_2\sin(\gamma\Omega s)\}}
\notag
\\
&= 
\int d^{n-2} \z \int_0^\infty dL \, L |\rho_{K}|^2 \, 
e^{-i \omega_{K} \gamma s}
\int_0^{2\pi}d\varphi \, e^{2i R L \sin(\gamma\Omega s/2) \sin(\varphi - \gamma\Omega s/2)}
\notag
\\
&= 
2\pi \int d^{n-2} \z \int_0^\infty dL \, L |\rho_{K}|^2 \, 
e^{-i \omega_{K} \gamma s}
\, J_0 \bigl(2 R L \sin(\gamma\Omega s/2) \bigr)
\notag
\\
&= 
2\pi \int d^{n-2} \z \int_0^\infty dL \, L |\rho_{K}|^2 \, 
e^{-i \omega_{K} \gamma s}
\sum_{m\in\BbbZ} J_m^2 (R L) \, e^{im\gamma\Omega s}
\ , 
\end{align}
where $K = \sqrt{L^2 + \z^2}$. 
At the second equality we have written 
$\k = (L\cos\varphi, L\sin\varphi, z_1, \ldots, z_{n-2})$, 
where $0\le L < \infty$ and $0 \le \varphi < 2\pi$.

We next write $\z$ in $(n-2)$-dimensional polar coordinates, 
so that $Z = |\z| \ge0$ is the radius. 
Integration over the $n-3$ angles gives the factor~$\Omega_{n-3}$, and we obtain 
\begin{align}
\mathcal{W}(s,0) 
&= 
\frac{4 \pi^{n/2}}{\Gamma \! \left(\frac{n-2}{2}\right)} 
\int\limits_{\substack{0\le Z < \infty\\
0 \le L < \infty}} dZ \, dL \, Z^{n-3} L |\rho_{K}|^2 \, 
e^{-i\omega_{K} \gamma s}
\sum_{m\in\BbbZ} J_m^2 (R L) \, e^{im\gamma\Omega s}
\notag
\\
&= 
\frac{4 \pi^{n/2}}{\Gamma \! \left(\frac{n-2}{2}\right)} 
\int_0^\infty dK \, K^{n-1} |\rho_{K}|^2 
\, e^{-i\omega_{K} \gamma s}
\notag
\\
& \hspace{6ex}
\times 
\sum_{m\in\BbbZ}
e^{im\gamma\Omega s}
\int_0^{\pi/2} d\alpha \, 
\sin\alpha \, {(\cos\alpha)}^{n-3}J_m^2 (R K \sin\alpha) 
\notag
\\
&= 
\frac{2}{{(4\pi)}^{n/2}}
\sum_{m\in\BbbZ}
\int_0^\infty dK \bigl| d(K/\M)\bigr|^2 
\frac{R^{2|m|} K^{2|m| + n-2}}{2^{|m|} \Gamma(|m|+1) \Gamma(|m|+n-\frac32)}
\notag
\\
& \hspace{12ex}
\times 
e^{-i(\omega_{K} - m \Omega)\gamma s}
\, 
{}_1 F_{2} \bigl( |m| + \tfrac12 ; 2|m|+1, |m|+n-\tfrac32 ; - R^2K^2\bigr) 
\ , 
\label{eq:helical-Wgen-final}
\end{align}
where ${}_1 F_{2}$ is the generalised hypergeometric function~\cite{dlmf}. 
The second equality in \eqref{eq:helical-Wgen-final} comes from writing 
$(Z,L) = (K \cos\alpha, K\sin\alpha)$ with $0\le\alpha\le \pi/2$. 
The third equality comes from using the identity 
\begin{align}
& \int_0^{\pi/2} d\alpha \, 
\sin\alpha \, {(\cos\alpha)}^{n-3}J_\mu^2 (z \sin\alpha) 
\notag
\\
& \hspace{4ex}
= 
\frac{z^{2\mu} \Gamma(\frac{n-2}{2})}{2^{2\mu+1} \Gamma(\mu+1) \Gamma(\mu+n-\frac32)}
\, 
{}_1 F_{2} \bigl( \mu + \tfrac12 ; 2\mu+1, \mu+n-\tfrac32 ; - z^2\bigr) 
\ , 
\label{eq:Jsquared-identity}
\end{align}
where $\mu> -1/2$, and the observation that $J_m^2(z) = J_{|m|}^2(z)$ for $m\in\BbbZ$. 
\eqref{eq:Jsquared-identity} may be verified by using 8.442.1, 3.621.5 and 8.384.1 
in \cite{grad-ryzh} and 5.5.5 in~\cite{dlmf}. 

Finally, substituting \eqref{eq:helical-Wgen-final} into~\eqref{eq:transrate-nosigma}, 
performing the integral over~$s$, using 
\eqref{eq:disprel-gen}
and~\eqref{eq:density-gen}, 
and writing $g = K/\M$, we have 
\begin{align}
\mathcal{F}(E) 
&= 
\frac{\M^{n-2}}{{(4\pi)}^{(n-2)/2} \gamma}
\sum_{m\in\BbbZ}
\frac{\Mtilde^{2|m|}}{2^{|m|} \Gamma(|m|+1) \Gamma(|m|+n-\frac32)}
\notag
\\
& \hspace{4ex}
\times 
\int_0^\infty dg \, |d(g)|^2 \, g^{2|m| + n -2}
\, 
{}_1 F_{2} \bigl( |m| + \tfrac12 ; 2|m|+1, |m|+n-\tfrac32 ; - \Mtilde^2 g^2\bigr) 
\notag
\\
& \hspace{9ex}
\times 
\delta \! \left(g f(g) - \frac{1}{\Mtilde} \left(mv - \frac{\Etilde}{\gamma}\right)\right)
\ , 
\end{align}
where $\Mtilde$, $\Etilde$ and $v$ are as in~\eqref{eq:Fhelical2+1-final}. 
Assuming that $gf'(g) + f(g)>0$ for $g>0$, 
we may perform the integral over~$g$, finding 
\begin{align}
\mathcal{F}(E) 
&= 
\frac{\M^{n-2}}{{(4\pi)}^{(n-2)/2} \gamma}
\sum_{m = \lceil \Etilde/(v\gamma)\rceil}^\infty 
\frac{\Mtilde^{2|m|} \, |{d(g_m)|}^2 \, 
g_m^{2|m| + n -2}}{2^{|m|} \Gamma(|m|+1) \Gamma(|m|+n-\frac32) 
\bigl( g_m f'(g_m) + f(g_m) \bigr)}
\notag
\\[1ex]
& \hspace{26ex}
\times 
{}_1 F_{2} \bigl( |m| + \tfrac12 ; 2|m|+1, |m|+n-\tfrac32 ; - \Mtilde^2 g_m^2\bigr) 
\ , 
\label{eq:n+1-helical-sumform}
\end{align}
where $g_m$ is the unique solution to~\eqref{eq:gm-def}. 

We would again be interested in $\mathcal{F}(E)$ in the limit $\Mtilde\to\infty$. 
The $(2+1)$-dimensional results of Section \ref{sec:helical2+1} suggest 
that when $v < f_c$, $\mathcal{F}(E)$ could be expected to 
tend in this limit to the usual massless scalar field transition rate, 
given now by 
\begin{align}
\Fcircnormal(E) 
&= 
\frac{1}{{(4\pi)}^{(n-2)/2} \gamma \, {R\vphantom{|}}^{n-2}}
\sum_{m = \lceil \Etilde/(v\gamma)\rceil}^\infty 
\frac{{(mv - \Etilde/\gamma)}^{2|m| + n - 2}}{2^{|m|+1} \Gamma(|m|+1) \Gamma(|m|+n-\frac32)}
\notag
\\[1ex]
& \hspace{16ex}
\times 
{}_1 F_{2} \bigl( |m| + \tfrac12 ; 2|m|+1, |m|+n-\tfrac32 ; - (mv - \Etilde/\gamma)^2 \bigr) 
\ , 
\label{eq:n+1-helical-unmod}
\end{align}
whereas when $v > f_c$, we could expect a large additional contribution, 
providing a low energy Lorentz violation. 
Unfortunately, we have not been able to investigate these expectations systematically. 
An analytic investigation would require a uniform asymptotic 
estimate for the summands in~\eqref{eq:n+1-helical-sumform}. 
A~numerical investigation would require an accurate numerical 
evaluation of the generalised hypergeometric functions 
in \eqref{eq:n+1-helical-sumform} and~\eqref{eq:n+1-helical-unmod}, 
including the large values of $m$ that could be expected to bring 
in the Lorentz-breaking contribution when $v > f_c$. 

Circular motion in four spacetime dimensions was 
investigated in~\cite{Stargen:2017xii}. 
Equation (25) therein agrees with the $n=3$ case of our 
\eqref{eq:n+1-helical-sumform} when $E\ge0$, 
and numerical evidence is presented from the regime 
$v < f_c$ and $E>0$, indicating that in this regime 
$\mathcal{F}(E) \to \Fcircnormal(E)$ as 
$\Mtilde\to\infty$. 
This numerical evidence is consistent with 
the expectations we have voiced above.

\section{Summary and discussion\label{sec:conclusions}} 

We have investigated the low energy phenomenology of a relativistic scalar field 
that violates Lorentz invariance at high energies, by coupling the field 
to an inertial Unruh-DeWitt detector in Minkowski spacetimes of dimension 
two or greater, and to an Unruh-DeWitt detector in uniform circular motion 
in Minkowski spacetimes of dimension three or greater. 
The dispersion relation was assumed to be Lorentz invariant at low 
energies but subluminal in some interval of high energies. 
We showed that the inertial detector experiences a large Lorentz violation 
at low energies in all spacetime dimensions greater than two when the 
detector's velocity in the preferred frame exceeds a critical velocity 
determined by the dispersion relation: this generalises results 
obtained previously in four dimensions in \cite{Kajuri:2015oza,Husain:2015tna}. 
We also showed that a similar large low energy Lorentz violation 
occurs for a detector in circular motion in $2+1$ dimensions. 
Finally, we laid the analytic groundwork for examining 
circular motion in all dimensions greater than $2+1$, 
generalising the $(3+1)$-dimensional analysis of~\cite{Stargen:2017xii}. 

On the theoretical side, our analysis was motivated by the high energy Lorentz 
violations that occur in many approaches to quantum gravity, 
and the hope to constrain these violations by low energy phenomenology. 
For example, as discussed in~\cite{Husain:2015tna}, 
the four-dimensional inertial motion results appear to experimentally rule out 
a field quantised in the polymer quantisation framework 
\cite{Ashtekar:2002sn,Ashtekar:2002vh,Husain:2010gb,Hossain:2010eb,Hossain:2009ru,Seahra:2012un} 
that is motivated by loop quantum gravity~\cite{Rovelli-book,Thiemann-book}, 
in the implementation of this framework adopted in~\cite{Hossain:2010eb}. 

On a more practical side, we anticipate that 
our results may be applicable to analogue spacetime laboratory experiments, 
where the effective spacetime dimension often differs from four
\cite{Belgiorno:2010wn,Weinfurtner:2010nu,Steinhauer:2015saa,Torres:2016iee,Leonhardt:2017lwm}.
We leave the development of this topic as a subject of future work. 

\section*{Acknowledgments}

We thank 
Antonin Coutant, 
Cisco Gooding, 
Viqar Husain, 
Sanjeev Seahra, 
Lakshmanan Sriramkumar, 
Silke Weinfurtner 
and 
L.~J. Zhou 
for helpful discussions and correspondence. 
JL was supported in part by Science and Technology Facilities Council
(Theory Consolidated Grants ST/J000388/1 and ST/P000703/1). 
SDU was supported by a 
summer research bursary from the School 
of Mathematical Sciences, University of Nottingham. 

\appendix

\section{Appendix: Proof of Theorem \ref{theorem:2+1:limittheorem}} 

In this appendix we give the proof of 
Theorem~\ref{theorem:2+1:limittheorem}. 

\subsection{Case (i): $0<v< f_c$} 

Suppose that $0<v< f_c$. 

From \eqref{eq:gm-def} we see that $\Mtilde g_m\to mv - \Etilde/\gamma$ as 
$\Mtilde\to\infty$, for each fixed~$m$. 
From 10.14.2 and 10.14.7 in \cite{dlmf} it follows 
that for sufficiently large $\Mtilde$ there exist 
constants $m_0\in\BbbZ_+$, $\alpha>0$ and $C>0$, independent of~$\Mtilde$, 
such that 
$0 < J_m(\Mtilde g_m) < C m^{-1/3} e^{-\alpha m}$
for all $m \ge m_0$. Since $|d(g)|$ and $\bigl(g f'(g) + f(g)\bigr)^{-1}$ 
are bounded under our assumptions, 
it follows by a dominated convergence argument that the $\Mtilde\to\infty$ 
limit in \eqref{eq:Fhelical2+1-sumform} can be taken under the sum. 
This completes the proof of Case~(i). 

\subsection{Case (ii): $f_c < v < 1$} 

Suppose that $f_c < v < 1$. Recall that under our assumptions the equation 
$f(g) = v$ then has exactly two solutions, denoted by 
$g_- \in (0,g_c)$ and $g_+ \in (g_c,\infty)$. 

To begin, we choose constants $g_1 \in (0, g_-)$ and $g_2 \in (g_+,\infty)$. 
Note that $v < f(g_1) < 1$ and $v < f(g_2)$.  For sufficiently large~$\Mtilde$, 
the set of integers $m$ for which $g_{m} \in (0,g_1)$ is nonempty and 
bounded from above, 
and we denote the largest integer in this set by~$m_1$. 
Similarly, the set of integers $m$ for which 
$g_m \in (g_2,\infty)$ is nonempty  
and bounded from below, 
and we denote the smallest integer in this set by~$m_2$.  
Note that $m_1$~and $m_2$ depend on~$\Mtilde$, 
they satisfy $m_1 < m_2$, and they both tend to infinity as $\Mtilde\to\infty$. 

Let $\Mtilde$ now be so large that $m_1 > 
\max \bigl(1, 1+ \lceil \Etilde/(v\gamma)\rceil\bigr)$. 
Let $\mathcal{F}_1(E)$, $\mathcal{F}_2(E)$ and $\mathcal{F}_3(E)$ be 
the contributions to $\mathcal{F}(E)$ \eqref{eq:Fhelical2+1-sumform} from respectively 
$m < m_1$, $m_1 \le m \le m_2$ and $m > m_2$. We need to estimate each of 
$\mathcal{F}_1(E)$, $\mathcal{F}_2(E)$ and $\mathcal{F}_3(E)$. 

In $\mathcal{F}_1(E)$ and $\mathcal{F}_3(E)$, 
10.14.2 and 10.14.7 in \cite{dlmf} again provide a dominated 
convergence bound that justifies taking the $\Mtilde\to\infty$ limit 
under the sum over~$m$. 
The outcome is that $\mathcal{F}_1(E) \to \Fcircnormal(E)$ 
and $\mathcal{F}_3(E) \to 0$
as $\Mtilde\to\infty$. 

In $\mathcal{F}_{2}(E)$, we view the sum over $m$ as the 
Riemann sum for an integral in the variable $m/\Mtilde$. 
Changing the integration variable from $m/\Mtilde$ to 
$g$ by \eqref{eq:gm-def} shows that the $\Mtilde\to\infty$ 
limit of $\mathcal{F}_{2}(E)$ 
is given by the $\Mtilde\to\infty$ limit of 
\begin{align}
\tilde{\mathcal{F}}_{\Mtilde}(E) 
&= 
\frac{1}{v\gamma}
\int_{g_1}^{g_2}
dg 
\, {|d(g)|}^2 
\, 
\Mtilde 
J^2_{\frac{\Mtilde}{v} \left(g f(g) + \frac{\Etilde}{\gamma\Mtilde}\right)}(\Mtilde g) 
\ , 
\label{eq:Fhelical2+1-int-integral}
\end{align}
provided the $\Mtilde\to\infty$ limit of the integrand
in \eqref{eq:Fhelical2+1-int-integral} exists and is sufficiently uniform in~$g$. 
That the limit has these properties under our assumptions 
follows from the uniform asymptotic expansion 
of $J_\nu(\nu z)$ as $\nu\to\infty$, given by 10.20.4 in~\cite{dlmf}, 
together with the asymptotic expansions of the Airy function $\Ai$ 
at large positive and negative argument, given by 9.7.5 and 9.7.9 in~\cite{dlmf}. 
The key property responsible for a nonvanishing answer in the limit is 
that 
\begin{align}
\nu J_\nu^2(\nu z) = \frac{2}{\pi \sqrt{z^2-1}}
\cos^2 \! \left[ \nu \! \left( \sqrt{z^2-1} - \arcsec z\right) - \frac{\pi}{4}\right]
+ o(1) 
\label{eq:nuJnuz-expansion}
\end{align}
as $\nu\to\infty$ with fixed $z>1$, 
together with an appropriate uniformity discussion in~\eqref{eq:nuJnuz-expansion}, 
and the use of the identity $2 \cos^2 \! z = 1 + \cos(2z)$ 
and the Riemann-Lebesgue lemma to justify 
the replacement $2 \cos^2 \to 1$ when \eqref{eq:nuJnuz-expansion} 
is used under the integral in~\eqref{eq:Fhelical2+1-int-integral}. 
We find that 
$\mathcal{F}_2(E) \to \Delta \mathcal{F}$ as $\Mtilde\to\infty$, 
where $\Delta \mathcal{F}$ is given by~\eqref{eq:DeltaFhelical2+1}. 

This completes the proof of Case~(ii). 
$\blacksquare$

\end{document}